\documentclass[a4paper,11pt]{article}
\usepackage{hyperref}
\title{On the security of 2-key triple DES}
\author{Chris J. Mitchell\\Information Security Group,
Royal Holloway, University of London\\
\url{www.chrismitchell.net}}
\date{19th February 2016 (typos fixed: 17th July 2016)}
\begin{document}

\bibliographystyle{plain}

\maketitle

\section*{Abstract}

This paper reconsiders the security offered by 2-key triple
DES, an encryption technique that remains widely used despite
recently being de-standardised by NIST\@.  A generalisation of
the 1990 van Oorschot-Wiener attack is described, constituting
the first advance in cryptanalysis of 2-key triple DES since
1990. We give further attack enhancements that together imply
that the widely used estimate that 2-key triple DES provides 80
bits of security can no longer be regarded as conservative; the
widely stated assertion that the scheme is secure as long as
the key is changed regularly is also challenged.  The main
conclusion is that, whilst not completely broken, the margin of
safety for 2-key triple DES is slim, and efforts to replace it,
at least with its 3-key variant, should be pursued with some
urgency.

\section{Introduction}

Despite the fact that it has long since been regarded as purely
of historical interest by many cryptographers, triple DES
remains of considerable practical importance, particularly in
the payments industry.  This is true of both its widely
discussed variants, i.e.\ 2-key and 3-key triple DES.

In late 2015, NIST finally withdrew support for 2-key triple
DES, something that had long been trailed and that does not
appear to have occurred because of any new insights into the
security of the scheme. However, this withdrawal of support
does not mean that the world has stopped using this variant,
and it also remains an ISO/IEC standard (albeit with ISO/IEC
having published warnings regarding the limited level of
security that it provides).

As discussed in the next section, the security of 2-key triple
DES has always been regarded as only giving a small margin of
safety.  In this paper we show that this margin is even less
than was previously thought.  We do this in three main ways:
\begin{itemize}
\item we show how the well-known van Oorschot-Wiener attack
    can be generalised to allow its effectiveness to be
    considerably improved by exploiting ciphertext
    generated using multiple keys;
\item building on the previous insight we show how the DES
    complementation property can be used to gain a factor
    of two efficiency improvement;
\item we demonstrate how partially known
    plaintext/ciphertext pairs can be used in the attack as
    well as fully known pairs, without significantly
    damaging the attack's computational or storage
    complexity.
\end{itemize}
We also briefly discuss possible practical approaches to the
implementation of attacks against 2-key triple DES, as well as
considering the impact of the generalised attack on the
security of the ANSI retail MAC.

As a result we conclude that the widely held assessment that
2-key triple DES only offers 80 bits of security is by no means
an overly conservative assumption.  It also follows that some
of the most significant advice given to users of 2-key triple
DES in order to help avoid cryptanalytic attacks is of limited
validity.

The remainder of the paper is organised as follows.  In
section~\ref{history} we briefly review the history of triple
DES\@. This is followed in section~\ref{van} by a discussion of
a generalisation of the van Oorschot-Wiener attack, which for
the last 25 years has been the most effective known attack
against 2-key triple DES\@. In section~\ref{complement} we then
show how the DES complementation property can be used to double
the attack speed.  We turn in section~\ref{partial} to
considering how partially known plaintext/ciphertext pairs can
be used in the attack.  We briefly review possible practical
attack implementation strategies in
section~\ref{implementation}, before discussing the use of the
attack method against the ANSI retail MAC in
section~\ref{ARMAC}.  The paper concludes in
section~\ref{conclusions}.

\section{Triple DES --- a brief history}  \label{history}

The DES block cipher was originally published as a US Federal
Standard (NBS FIPS PUB 46 \cite{kn:108}) as long ago as 1977
--- for further details of its origins see, for example,
chapter 7 of Menezes, van Oorschot and Vanstone \cite{K276}.
DES, also known as the \emph{Data Encryption Algorithm (DEA)},
is a 64-bit block cipher, i.e.\ it transforms a 64-bit
plaintext block into a 64-bit ciphertext block, employing a
56-bit key.

From the moment it was published it was criticised for the
short length of its key. Even 40 years ago, performing $2^{56}$
encryption operations, as necessary to perform a brute force
search for the key using a single known plaintext/ciphertext
pair, was just about within the bounds of possibility using
special-purpose hardware. Indeed, the design for a special
purpose brute-force DES-breaking machine capable of finding a
DES key within a day was sketched by Diffie and Hellman,
\cite{K688}, and was estimated by them to cost 10 million US
dollars.

These concerns did not stop the very widespread adoption of
DES, not only within the US (where it became an ANSI standard,
X3.92 \cite{kn:107}), but worldwide, particularly within the
financial sector. This was probably because of the lack of any
widely known public competitor schemes. Despite the short key
length, the use of DES was arguably a huge success and there
are no public domain examples, of which the author is aware, of
significant compromises because of the limited key length.

However, it became clear within a few years of its publication
that a more secure version of DES was required, allowing a
longer key length. This need gave rise to the two well-known
versions of triple DES, one of which forms the main focus of
this paper. With the rise of triple DES, use of just one
iteration of DES, as originally standardised, became known as
single DES.

The gradual switch to triple DES was supported by its
standardisation by NIST \cite{K363,K767}, ANSI \cite{K769}, and
ISO/IEC \cite{K534,K702}. This switch was very timely, as by
the second half of the 1990s single DES has been broken in
various ways (all using brute force attacks). Of particular
interest was the fact that in 1998 a special-purpose DES
breaking machine, Deep Crack, was designed and built for a few
hundreds of thousands of US dollars \cite{K770}, vindicating
the 1977 predictions by Diffie and Hellman. In fact, a large
distributed software-only attack (coordinated through the
\emph{DESCHALL project}) had shortly before succeeded in
breaking a `DES challenge' \cite{K774}. This served notice to
the world at large that single DES was no longer secure.

Since double DES (i.e.\ two iterations of DES encryption using
independent keys) has long been ruled out as offering very
limited additional security by comparison with single DES (see
Diffie and Hellman, \cite{K688}), then the obvious next
alternative is to perform triple DES, i.e.\ three iterations of
the DES algorithm. The general idea of using three iterations
of DES was mentioned in 1977 by Diffie and Hellman,
\cite{K688}, as a way of dramatically improving the security of
DES\@. In practice, rather than performing three consecutive
encryptions, it has become the norm to first perform an
encryption (using key $K_1$), then perform a decryption (using
key $K_2$), and finally perform another encryption (using key
$K_3$). The encrypt-decrypt-encrypt approach has the advantage
of being backwards-compatible with single DES if $K_1=K_2=K_3$.
This potentially makes migration from single to triple DES much
simpler. If $K_1$, $K_2$ and $K_3$ are all chosen
independently, encrypt-decrypt-encrypt is known as 3-key triple
DES.

As reported by Merkle and Hellman, \cite{K771}, in 1978 Tuchman
proposed a 2-key variant of triple DES. This involves choosing
$K_1=K_3$, i.e.\ first encrypting with $K_1$, then decrypting
with $K_2$, and finally re-encrypting with $K_1$. This approach
has the advantage of only involving two DES keys, reducing the
key storage and transmission requirements to the same as for
double DES, but giving significantly greater security than
provided by double DES\@. However, it is clear that 2-key
triple DES is itself significantly less secure than 3-key
triple DES, and Merkle and Hellman \cite{K771} described an
attack against 2-key triple DES which is significantly more
effective than the best known attack against the 3-key version.
They suggested that this means that the 3-key variant should
always be used. Since Merkle and Hellman's attack was published
in 1981, one other attack against 2-key triple DES has been
devised, namely that due to van Oorschot and Wiener,
\cite{K350}; this latter attack is discussed in
section~\ref{van} below.

The NIST standard for the DES algorithm, FIPS PUB 46-3
\cite{K363}, was withdrawn back in 1999. This signalled the end
of standard-status for single DES. The situation for 2-key and
3-key triple DES standardisation is much less clear cut. Triple
DES has been standardised by a variety of bodies including NIST
in SP 800-67, \cite{K767}, and by ISO/IEC in the first and
second editions of ISO/IEC 18033-3, \cite{K534,K702}. All these
standards specify both 2-key and 3-key triple DES.

Despite the fact that 2-key triple DES is clearly less secure
than the 3-key version, it has been very widely used,
particularly by the electronics payments industry, where it
remains in active use. For example, the current version of the
EMV standard, \cite{K777}, used as the basis for security for
credit and debit cards worldwide, specifies that `The
double-length key triple DES encipherment algorithm (see
ISO/IEC 18033-3) is the approved cryptographic algorithm to be
used in the encipherment and MAC mechanisms' [here double
length is a reference to the 2-key variant of triple DES].

As a result, there is considerable industry pressure to retain
both variants as standards. At the same time, there has been
considerable pressure both from academia and from bodies such
as NIST to phase out all use of DES (and in particular 2-key
triple DES) in favour of more modern, more secure, and more
efficient algorithms, such as AES, \cite{K361}.

In this latter connection, for some years NIST has been
particularly keen to phase out triple DES, particularly the
2-key variant. Indeed, in the latest revision of NIST SP
800-131A, \cite{NIST800-131A}, published in late 2015, it was
announced that support for 2-key triple DES had been withdrawn.
A similar statement can be found in the latest (January 2016)
version of NIST SP 800-57 Part 1, \cite{K776}.  This withdrawal
of support is in line with previous announcements on the
subject. However, ISO/IEC has not followed the same path, and
both 2-key and 3-key triple DES variants remain as standard
algorithms in the most recent (2010) version of ISO/IEC 18033-3
\cite{K702}, although use of the 3-key variant is recommended.
ISO/IEC SC 27 (the committee responsible for drawing up ISO/IEC
18033-3) has published guidance on the use of triple DES in two
standing documents, \cite{K772,K773}.  Key statements from one
of these standing documents, \cite{K772}, expressing sentiments
that have been widely reproduced elsewhere, are:
\begin{itemize}
\item `depending on the required security level, the
    maximum number of plaintexts encrypted under a single
    key should be limited'; and
\item `the effective key-length of two-key Triple-DES in
    specific applications can only be regarded as 80 bits
    (instead of 112 bits)'.
\end{itemize}
The statement regarding 80-bit security has also been given in
various documents produced by NIST (see, for example, Section
5.6.1 of NIST 800-57 Part 1, \cite{K776}).  We reconsider both
these claims at the end of this paper.

\section{Generalising the van Oorschot-Wiener attack}
\label{van}

\subsection{The original attack}

In 1990, almost a decade after the Merkle-Hellman attack was
published, a somewhat more practical attack against 2-key
triple DES was described by van Oorschot and Wiener
\cite{K350}. This attack is more practical than Merkle-Hellman
in that it only requires known plaintext/ciphertext pairs,
rather than chosen plaintext/ciphertext pairs.  We next provide
a brief description of this attack.

The attack requires that the attacker has access to $n$
plaintext/ciphertext pairs ($P$, $C$), all created using the
same 2-key triple DES key (i.e.\ the same pair of DES keys
($K_1$, $K_2$) --- we use this notation throughout).  The main
idea behind the attack is to fix a 64-bit value $A$, and to
hope that $e_{K_1}(P)=A$ for one of the known pairs ($P$, $C$).
If this is true, then finding $K_2$ only requires a single DES
key search, i.e.\ performing $2^{56}$ DES operations.  Of
course, unless $n$ is very large, the guess is unlikely to be
true, so the attack has to be performed for many values of $A$.
The larger the value of $n$, then the larger the probability of
a successful guess of a value $A$, and hence the more efficient
the attack.

The attack proceeds as follows, where $e_K(P)$ and $d_K(C)$
represent the DES-encryption of $P$, and DES-decryption of $C$,
respectively, using the key $K$.
\begin{enumerate}
\item Tabulate the ($P$, $C$) pairs, sorted or hashed on
    the plaintext values $P$, to create Table 1, which
    requires $O(n)$ words of storage.
\item Now randomly select and fix (for steps 2--4) a value
    $A$. [This stage of the attack will succeed if and only
    if $A = e_{K_1}(P)$ for one of the known plaintexts
    $P$. If steps 2-4 succeed with this value of $A$ we can
    find the target triple DES key; if not, we simply
    repeat with a different value of $A$ --- see step 5.]
\item Create a second table (Table 2) as follows. For each
    of the $2^{56}$ possible DES keys $i$, calculate $P_i =
    d_i(A)$. Next look up $P_i$ in Table 1. If $P_i$ is
    found in the first column of Table 1, take the
    corresponding ciphertext value $C$ and compute $B =
    d_i(C)$. Now store $B$ together with $i$ in Table 2,
    which is sorted (or hashed) on the $B$ values.  Note
    that the same $B$ value may occur more than once.
\item Each entry in Table 2 consists of a value of $B$ and
    the corresponding key $i$, where $i$ is a candidate for
    $K_1$; as described above, each ($B$, $i$) pair is
    associated with a ($P$, $C$) pair from Table 1 where
    $e_i(P) = A$. The remaining task is to search for
    possible values of $K_2$.

    For each of the $2^{56}$ candidates, $j$, for $K_2$,
    calculate what the value $B$ would be if $j$ had been
    used for $K_2$, i.e.\ $B_j = d_j(A)$. Now look up $B_j$
    in Table 2. For each appearance of $B_j$ (if any) the
    corresponding key $i$ from Table 2, along with key $j$,
    is a candidate for the desired pair of keys ($K_1$,
    $K_2$). Each such candidate key pair is then tested on
    at most two other plaintext/ciphertext pairs. If this
    key pair gives the correct results then the target
    triple DES key ($K_1$, $K_2$) has been found and the
    task is complete.
\item If the algorithm does not succeed, then the process
    in steps 2--4 is repeated for a new random value of
    $A$. [Note that, to avoid the (small) risk of repeating
    values of $A$, the values could be worked through in
    some order].
\end{enumerate}

We can summarise the complexity of this attack as follows.
\begin{itemize}
\item The time required to create and sort/hash Table 1 is
    negligible compared to other computations given $n <<
    2^{56}$. As already mentioned the space required is
    $O(n)$.
\item For each trial value $A$, Table 2 costs a little more
    than $2^{56}$ DES computations to create (assuming
    Table 1 is hashed on the plaintext values so that
    look-ups take a constant time). Because only $2^{56}$
    out of $2^{64}$ possible 64-bit blocks are searched for
    in Table 1, the expected number of entries in Table 1
    is $n/2^8$, i.e.\ the storage required for Table 2 is
    negligible by comparison with Table 1.

\item Working with Table 2 to find candidate pairs of keys
    costs a further $2^{56}$ DES computations.  That is,
    testing a single value of $A$ costs a total of around
    $2^{57}$ DES computations.

\item The probability that a single iteration of steps 2--4
    will succeed, i.e.\ yield the correct key pair, is
    approximately $n/2^{64}$, and hence the total cost of
    the attack is approximately $2^{121}/n$ DES
    computations (assuming the cost of the various look-ups
    and tests is dwarfed by the DES calculations).
\end{itemize}
In summary, if we have $2^t$ known plaintext/ciphertext pairs,
i.e.\ $n=2^t$, then 2-key triple DES can be broken using
$2^{121-t}$ DES computations and $O(2^t)$ storage. For example,
if $n=2^{32}$, i.e.\ if we have as many as 4 billion known
plaintext/ciphertext pairs, then the key can be discovered in
$2^{89}$ DES computations.

The conclusion from the above attack is that launching a
practical attack only becomes practical if very large volumes
of matching plaintext and ciphertext, all generated using a
single triple DES key, are available. This has led to the
following two widely drawn conclusions regarding the security
of 2-key triple DES, referred to at the end of
section~\ref{history}.
\begin{itemize}
\item As a conservative estimate, 2-key triple DES offers
    at least 80 bits of security.
\item As long as the key is changed reasonably frequently
    (limiting $n$ in the above attack), practical attacks
    against 2-key triple DES remain infeasible.
\end{itemize}
In the remainder of this paper we challenge the second
conclusion, and also provide evidence that the lower bound
estimate of 80 bits of security is not as conservative as it
might seem.

\subsection{The generalisation}  \label{generalised}

We start by making an apparently simple observation on the van
Oorschot-Wiener attack.  That is, the attack will work
\emph{just as well} if the $n$ plaintext/ciphertext pairs are
generated using a range of different triple DES keys.  Of
course, when performing the tests in step 5, it is necessary to
use additional plaintext/ciphertext pairs that have been
generated using the appropriate triple DES key.  Also, when the
attack is successful, only one of the keys will be found, and
can only be used to decrypt other material encrypted using that
key. Nevertheless, depending on the application, this could
still have devastating consequences for security.

We can modify the algorithm described above to take account of
this observation by changing steps 1, 3 and 4, as follows.
\begin{enumerate}
\item[$1'$] Assemble the pairs ($P$, $C$) into
    subsets\footnote{We require that each subset contains
    at least two, and preferably three, pairs, so that
    candidates for ($K_1$, $K_2$) can be checked.}, where
    all the pairs in each subset have been created using
    the same key, and assign each subset a unique label $s$
    (we also use $s$ as the label for the triple DES key
    used to create the subset). Tabulate all the ($P$, $C$,
    $s$) triples (where $s$ is the key label), sorted or
    hashed on the plaintext values, to create Table 1,
    which requires $O(n)$ words of storage. Note that there
    may be repeated $P$ values, but this should not create
    a major implementation difficulty.
\item[$3'$] Create a second table (Table 2) as follows. For
    each of the $2^{56}$ possible DES keys $i$, calculate
    $P_i = d_i(A)$. Next look up $P_i$ in Table 1. If $P_i$
    is found in the first column of Table 1, take the
    corresponding ciphertext value (or values) $C$ (and the
    label(s) $s$) and for each compute $B = d_i(C)$. Now
    store $B$ together with $i$ and $s$ in Table 2, which
    is sorted (or hashed) on the $B$ values.  Note that the
    same $B$ value may occur more than once.
\item[$4'$] Each entry in Table 2 consists of a value of
    $B$ and the corresponding key $i$ and label $s$, where
    $i$ is a candidate for $K_1$ for label $s$; as
    described above, each ($B$, $i$, $s$) triple is
    associated with a ($P$, $C$, $s$) triple from Table 1
    where $e_i(P) = A$. The remaining task is to search for
    possible values for $K_2$.

    For each of the $2^{56}$ candidates, $j$, for $K_2$,
    calculate what the value $B$ would be if $j$ had been
    used for $K_2$, i.e.\ $B_j = d_j(A)$. Now look up $B_j$
    in Table 2. For each appearance of $B_j$ (if any) the
    corresponding key $i$ from Table 2, along with key $j$,
    is a candidate for the desired pair of keys ($K_1$,
    $K_2$) with label $s$. Each such candidate key pair is
    then tested on at most two other plaintext/ciphertext
    pairs from the label $s$ subset. If this key pair gives
    the correct results then the triple DES key ($K_1$,
    $K_2$) with label $s$ has been found and the task is
    complete.
\end{enumerate}

Apart from the fact that the tables have an additional value in
each entry (namely the key label, which might typically be at
most four bytes long), none of the attack complexities have
changed. I.e., if we have $2^t$ known plaintext/ciphertext
pairs, i.e.\ $n=2^t$, then 2-key triple DES can be broken using
$2^{121-t}$ DES computations and $O(2^t)$ storage.  Here the
meaning of `broken' is slightly different from previously, in
that it means that one of the triple DES keys has been
discovered, rather than the single key used to encrypt the
entire set of $n$ pairs.

It is important to see that this means that changing the triple
DES key from time to time has no impact on the effectiveness of
the attack.  Of course, regular key changes remain a good idea
since, even if the above attack is successful, only the
plaintext encrypted using the broken key can be recovered.
Finally, we observe that there is nothing specific to DES about
the above generalisation, or the original van Oorschot-Wiener
attack for that matter.  The attack would work equally well
against any triple-iterated block cipher with the same key
structure; however, we restrict our attention to DES here since
it is the only block cipher for which triple encryption is
widely used (at least as far as is known to the author). Also,
the enhancement described in the next section \emph{is}
specific to DES.

\section{Exploiting the DES complementation property}
\label{complement}

We next see how the effectiveness of the generalised van
Oorschot-Wiener attack can be improved using the well-known DES
complementation property (see, for example, \cite{K276}). This
property says that, for any 64-bit block $P$ and any DES key
$K$ ( where $\overline{X}$ denotes the bit-wise complement of
bit string $X$):
\[ \overline{e_K(P)} = e_{\overline{K}}(\overline{P}). \]
That is, if $P$ and $K$ are complemented, then the output
ciphertext is also complemented.  It is interesting to observe
that Lucks \cite{K780} considered how to use this property to
improve the efficiency of his attacks on 3-key triple DES.

This property can be used to double the number of
plaintext/ciphertext pairs available to conduct the attack,
since every plaintext/ciphertext pair for the key $K$ will give
us another pair for the key $\overline{K}$.  Another way of
looking at this is that we can perform the attack steps for $A$
and $\overline{A}$ simultaneously.

We can incorporate this observation into the generalised attack
of section~\ref{van} by modifying steps $3'$ and $4'$ as
follows.
\begin{enumerate}
\item[$3''$]  Create a second table (Table 2) as follows.
    For each of the $2^{56}$ possible DES keys $i$,
    calculate $P_i = d_i(A)$. Next look up both $P_i$ and
    $\overline{P_i}$ in Table 1.
    \begin{itemize}
    \item If $P_i$ is found in the first column of
        Table 1, take the corresponding ciphertext
        value (or values) $C$ (and the label(s) $s$)
        and for each compute $B = d_i(C)$. Now store
        $B$ together with $i$, $s$ and a one-bit
        complementation flag $F$ (in this case set to
        zero) in Table 2, which is sorted (or hashed)
        on the $B$ values.
    \item Similarly, if $\overline{P_i}$ is found in
        the first column of Table 1, take the
        corresponding ciphertext value (or values) $C$
        (and the label(s) $s$ and flag $F=1$ to
        indicate a complemented value) and for each
        compute $B = d_i(C)$. Now store $\overline{B}$
        together with $\overline{i}$, $s$ and $F$ in
        Table 2, which is sorted (or hashed) on the $B$
        values.
    \end{itemize}
Note that the same $B$ value may occur more than once. Note
also that, instead of introducing the flag $F$, we could
choose to implement Table 2 in two parts, one containing
the entries with $F=0$ and the other the entries with
$F=1$. Such an approach might simplify the implementation
of step $4''$.
\item[$4''$] Each entry in Table 2 consists of a value of
    $B$ and the corresponding key $i$, label $s$ and flag
    $F$, where $i$ is a candidate for $K_1$ for label $s$;
    as described above, each ($B$, $i$, $s$) triple is
    associated with a ($P$, $C$, $s$) triple from Table 1
    where $e_i(P) = A$ if $F=0$ and $e_i(P) = \overline{A}$
    if $F=1$.

    The remaining task is to search for the desired value
    of $K_2$. For each of the $2^{56}$ candidates, $j$, for
    $K_2$, calculate what the value $B$ would be if $j$ had
    been used for $K_2$, i.e.\ $B_j = d_j(A)$ (note also
    that $\overline{B_j}=d_{\overline{j}}(\overline{A})$).
    Now look up $B_j$ and $\overline{B_j}$ in Table 2.
    \begin{itemize}
    \item For each appearance of $B_j$ with $F=0$ (if
        any) the corresponding key $i$ from Table 2,
        along with key $j$, is a candidate for the
        desired pair of keys ($K_1$, $K_2$) with label
        $s$. Each such candidate key pair is then
        tested on at most two other
        plaintext/ciphertext pairs from the label $s$
        subset. If this key pair gives the correct
        results then the triple DES key ($K_1$, $K_2$)
        with label $s$ has been found and the task is
        complete.
    \item  Similarly, for each appearance of
        $\overline{B_j}$ with $F=1$ (if any) the
        corresponding key $\overline{i}$ from Table 2,
        along with key $\overline{j}$, is a candidate
        for the desired pair of keys ($K_1$, $K_2$)
        with label $s$. Each such candidate key pair is
        then tested on at most two other
        plaintext/ciphertext pairs from the label $s$
        subset. If this key pair gives the correct
        results then the triple DES key ($K_1$, $K_2$)
        with label $s$ has been found and the task is
        complete.
    \end{itemize}
\end{enumerate}

Since, in effect, two values of $A$ are being tested at once,
the above modification should halve the number of times the
process needs to be performed.  At the same time, Table 2 will
contain twice as many entries, but since Table 2 is small by
comparison with Table 1, this should not significantly affect
the overall storage complexity.

Hence, if we have $2^t$ known plaintext/ciphertext pairs, i.e.\
$n=2^t$, then 2-key triple DES can be broken using $2^{120-t}$
DES computations and $O(2^t)$ storage. For example, if
$n=2^{32}$, i.e.\ if we have as many as 4 billion known
plaintext/ciphertext pairs, then the key can be discovered in
$2^{88}$ DES computations.

\section{Using partially known plaintext}  \label{partial}

The next modification to the attack that we describe is
designed to cope with the situation where we have ciphertext
blocks for which we do not know the precise plaintext value.
For example, we may have a ciphertext block $C$ for which we
know 56 of the 64 plaintext bits, but not the other eight,
i.e.\ there is a set of $2^8$ possible values for $P$ for a
given ciphertext block $C$. The van Oorschot-Wiener attack (and
the variants we have so far described) cannot use such
information, rather restricting the scenarios in which the
attack will work.

Such a situation could easily arise in practice.  To take a
simple example from the payments industry (where 2-key triple
DES is in use), the ISO 9564-1 \cite{K775} Format 0 PIN block
involves creating a 64-bit plaintext block by combining an
account number with a 4-digit PIN \cite{K775}.  If a
triple-DES-enciphered Format 0 PIN block is obtained for which
the account number is known, then the only unknown information
in the plaintext is the value of the PIN, for which there are
only $10^4\approx2^{13}$ possible values.

Such partial plaintext information can be used in a further
modification to the van Oorschot-Wiener attack. This
modification arises from the observation that the attack will
still work even if some of the plaintext/ciphertext pairs are
actually false. If a false pair generates a candidate key, then
this key will be rejected when it is checked against `correct'
pairs.  Of course, there is the danger that the check at the
end of step 4 might be done using a false pair, and hence a
valid candidate would be rejected, but we can avoid this if we
assume checking is always done using valid data.

This observation can be used to make use of partial knowledge
of a plaintext block (for a known ciphertext block) by
generating a set of plaintext/ciphertext pairs all having the
same ciphertext element. That is, we generate all the plaintext
blocks $P$ which satisfy the known information, and for each
such `possible' plaintext block we create a pair containing it
and the known ciphertext block.  We then add them all to the
set of known plaintext/ciphertext pairs used in the attack. For
example, if we have a ciphertext block $C$ for which we know
all but $w$ bits of the plaintext block, we then generate $2^w$
plaintext/ciphertext pairs with plaintext blocks covering all
possibilities for the `missing' $w$ bits, all with the same
value of $C$. Of course, all but one of these pairs will be
false, but this does not matter.

To see how this affects the attack, we give below a modified
version of the generalised attack technique given in
section~\ref{generalised} --- only steps 1 and 4 are changed,
and hence we only show these steps. Whilst we could readily
combine this modification with the attack exploiting the
complementation property, in order to simplify the presentation
we avoid doing this here.

We suppose that we start with $n$ ciphertext values, for some
of which we know the correct plaintext and for others we only
have partial information. We assume that in every case there
are at most $2^w$ candidates for the plaintext block, i.e.\ the
set of mostly false pairs for a single ciphertext block
contains at most $2^w$ pairs.

\begin{enumerate}
\item[$1'''$] Assemble the pairs ($P$, $C$) into subsets
    including the sets of mostly false pairs (as above),
    where all the pairs in each subset have been created
    using the same key, and assign each subset a label $s$.
    Note that there will be at most $2^wn$ pairs. Tabulate
    all the ($P$, $C$, $s$) triples, sorted or hashed on
    the plaintext values, to create Table 1, which requires
    $2^wO(n)$ words of storage. Note that there may be
    repeated $P$ values, but this should not create a major
    implementation difficulty.
\item[$4'''$] Each entry in Table 2 consists of a value of
    $B$ and the corresponding key $i$ and label $s$, where
    $i$ is a candidate for $K_1$ for label $s$; as
    described above, each ($B$, $i$, $s$) triple is
    associated with a ($P$, $C$, $s$) triple from Table 1
    where $e_i(P) = A$. The remaining task is to search for
    possible values for $K_2$.

    For each of the $2^{56}$ candidates, $j$, for $K_2$,
    calculate what the value $B$ would be if $j$ had been
    used for $K_2$, i.e.\ $B_j = d_j(A)$. Now look up $B_j$
    in Table 2. For each appearance of $B_j$ (if any) the
    corresponding key $i$ from Table 2, along with key $j$,
    is a candidate for the desired pair of keys ($K_1$,
    $K_2$) with label $s$. Each such candidate key pair is
    then tested on at most two other plaintext/ciphertext
    pairs from the label $s$ subset (where either the
    mostly false pairs are avoided, or where only the
    partial information about the plaintext is used in the
    checking). If this key pair gives the correct results
    then the triple DES key ($K_1$, $K_2$) with label $s$
    has been found and the task is complete.
\end{enumerate}

It remains for us to consider the complexity of this modified
attack.
\begin{itemize}
\item Table 1 will contain at most $2^wn$ entries.  The
    time required to create and sort/hash Table 1 remains
    negligible compared to other computations as long as $n
    << 2^{56-w}$. The space required is $2^wO(n)$.
\item For each trial value $A$, Table 2 costs a little more
    than $2^{56}$ DES computations to create (assuming
    Table 1 is hashed on the plaintext values so that
    look-ups take a constant time). Because only $2^{56}$
    out of $2^{64}$ possible 64-bit blocks are searched for
    in Table 1, the expected number of entries in Table 1
    is $2^{w-8}n$, i.e.\ the storage required for Table 2
    is negligible by comparison with Table 1.

\item Working with Table 2 to find candidate pairs of keys
    costs a further $2^{56}$ DES computations.  That is,
    testing a single value of $A$ costs a total of around
    $2^{57}$ DES computations.

\item The probability of a single iteration of steps 2--4
    succeeding, i.e.\ yielding the correct key pair, is
    approximately $n/2^{64}$, and hence the total cost of
    the attack is approximately $2^{121}/n$ DES
    computations (assuming the cost of the various look-ups
    and tests is dwarfed by the DES calculations).
\end{itemize}
In summary, if we have $2^t$ partially known
plaintext/ciphertext pairs, i.e.\ $n=2^t$, and we assume $n <<
2^{56-w}$, then 2-key triple DES can be broken using
$2^{121-t}$ DES computations and $O(2^{t+w})$ storage. For
example, if $n=2^{32}$, i.e.\ if we have as many as 4 billion
known (or partially known) plaintext/ciphertext pairs, then the
key can be discovered in $2^{89}$ DES computations. That is,
the extra work introduced through the use of `false' pairs is
minimal as long as $n << 2^{56-w}$, i.e.\ $t+w << 56$.  Of
course, the cost of storage has increased to $O(2^{t+w})$, but
this is still relatively modest if $t+w << 56$.  Note that we
can reduce the total number of DES computations to $2^{120-t}$
by combining the above modification with that given in
section~\ref{complement}.

Returning to the PIN block example above (for which
$w\approx13$), if $n=2^{32}$ then the attack complexity would
not be significantly different to the case where $2^{32}$ fully
known plaintext blocks are available.

In summary we have generalised the attack to the case where
only partial known plaintext is available, without
significantly increasing the attack complexity.  This, while
not simplifying the attack, means it will potentially apply in
many more practical scenarios.

\section{Implementation strategies}  \label{implementation}

Whilst performing an attack on 2-key triple DES will clearly be
a non-trivial computation, it is perhaps worth considering how
it might actually be done in practice.  Note that while we
refer to steps 1--5 from the unmodified van Oorschot attack,
the remarks below also apply to all the modified versions
described above.

First note that step 1 is a one-off computation working with
the known plaintext-ciphertext material to create Table 1.
This step should be performed carefully to optimise the cost of
the look-ups performed using Table 1 in subsequent parts of the
attack.

We next observe that there are obvious ways in which the
remainder of the attack can be parallelised.
\begin{itemize}
\item Performing steps 2--4 for a particular value of $A$
    is completely independent of performing them again for
    a different value of $A$.  All that is required is
    access to a copy of Table 1, generated by step 1. That
    is, software could be created which generated random
    values of $A$ and performed steps 2--4, and this
    software could be run without reference to other
    running copies of the software --- the only requirement
    is an effective random number generator so that
    different instances of the software generate different
    values of $A$ (with high probability).
\item The creation of Table 2 in Step 3 could be
    parallelised by partitioning the set of possible keys
    $i$, so that multiple machines together create Table 2.
\item Similarly, the use of Table 2 in step 4 could be
    partitioned by partitioning the set of possible keys
    $j$.  Note that each device performing this part of the
    attack will require a copy of Table 2.  Note also that,
    as they are found, `candidate' keys could be sent to a
    different device for testing using entries from Table
    1.
\end{itemize}

\section{Attacking the ANSI Retail MAC}  \label{ARMAC}

\subsection{Background}

As a slight digression we also consider the impact of the van
Oorschot-Wiener attack on the ANSI Retail Message
Authentication Code (MAC) \cite{K128}.  This MAC algorithm
appears to be used in the payments industry, since it is
standardised in A1.2.1 of the current version of EMV Book 2
\cite{K777}. The scheme, otherwise known as CBC-MAC-Y or
ISO/IEC 9797-1 algorithm 3 \cite{K313}, operates as follows.
For the purposes of this paper we describe it in the context of
use with DES, although the remarks apply more generally.  We
also use the same notation as employed previously.

A message $D$ to be MAC-protected is first padded and split
into a sequence of $q$ $n$-bit blocks: $D_1, D_2, \ldots,
D_q$\@. The MAC scheme uses a pair of keys $K_1$, $K_2$. The
MAC computation is as follows.
\begin{eqnarray*}
H_1 & = & e_{K_1}(D_1),\\
H_\ell & = & e_{K_1}(D_\ell\oplus H_{\ell-1}),\:\:\:(2\leq \ell\leq q), \mbox{ and }\\
M & = & e_{K_1}(d_{K_2}(H_q)),
\end{eqnarray*}
where $\oplus$ represents bit-wise exclusive or, and $M$ is the
MAC\@. Note that, for simplicity, we assume that the MAC is not
truncated.

It is not hard to see that this amounts to encrypting the
message using single DES in CBC mode, but using 2-key triple
DES on the final block; the MAC $M$ is then simply the
encryption of the final block.  This suggests that the van
Oorschot-Wiener attack may be relevant (and it is!).

The most effective general-purpose key recovery attack on the
ANSI retail MAC algorithm requires $2^{57}$ DES operations and
$2^{32}$ known message/MAC pairs, as described by Preneel and
van Oorschot \cite{K300}. An alternative key recovery attack,
requiring only one known MAC/message pair but a larger number
of verifications, is due to Knudsen and Preneel, \cite{K292};
this attack requires $2^{56}$ DES operations, one known
message/MAC pair, and $2^{56}$ online MAC verifications.
Further key recovery attacks based on MAC verifications have
been devised, \cite{K358,K364}, although they are more relevant
in the case where the MAC is truncated and so we do not
describe them further here.

\subsection{Applying the van Oorschot-Wiener attack}

First observe that the applicability of the van Oorschot-Wiener
attack to the ANSI retail MAC does not appear to have
previously been considered, very probably because the
`standard' Preneel-van Oorschot attack is typically more
effective.  As discussed in section~\ref{van}, the van
Oorschot-Wiener attack requires large volumes of matching
plaintext and ciphertext generated using a single key in order
to be effective. Also, as discussed immediately above, the
Preneel-van Oorschot attack, \cite{K300}, requires $2^{32}$
known message/MAC pairs, and if such material is available it
is then significantly more efficient than the van
Oorschot-Wiener attack.  That is, the van Oorschot-Wiener
attack does not appear to offer any advantage over the
established Preneel-van Oorschot attack.

However, the fact that van Oorschot-Wiener can be made to work
where the known ciphertext has bene generated using multiple
keys, suggests that it may have significance to this MAC
scheme.  We next sketch how the attack can be applied in this
case.  For simplicity we look at the application of the
`standard' version of the attack (as described in
section~\ref{van}), although the generalised versions of
sections~\ref{generalised} and \ref{complement} also apply.

We suppose the attacker has access to $n$ message/MAC pairs
(($D_1, D_2, \ldots, D_q$), $M$), all created using the same
pair of DES keys ($K_1$, $K_2$).  Note that, for simplicity, we
consider the padded and split version of a message. As before,
we fix a 64-bit value $A$, and in this case hope that
$d_{K_1}(M)=A$ for one of the known pairs (($D_1, D_2, \ldots,
D_q$), $M$). If this is true, then finding $K_2$ only requires
a single DES key search, i.e.\ performing $2^{56}$ DES
operations.  Of course, unless $n$ is very large, the guess is
unlikely to be true, so the attack has to be performed for many
values of $A$. The larger the value of $n$, then the larger the
probability of a successful guess of a value $A$, and hence the
more efficient the attack.

The attack proceeds as follows.
\begin{enumerate}
\item Tabulate the (($D_1, D_2, \ldots, D_q$), $M$) pairs,
    sorted or hashed on the values of $M$, to create Table
    1, which requires $O(2^rn)$ words of storage if we make
    the simplifying assumption that $q$ is bounded above by
    $2^r$.
\item Now randomly select and fix (for steps 2--4) a value
    $A$. [This stage of the attack will succeed if and only
    if $A = d_{K_1}(M)$ for one of the known values of $M$.
    If steps 2--4 succeed with this value of $A$ we can
    find the target key pair ($K_1$, $K_2$); if not, we
    simply repeat with a different value of $A$ --- see
    step 5.]
\item Create a second table (Table 2) as follows. For each
    of the $2^{56}$ possible DES keys $i$, calculate $M_i =
    e_i(A)$. Next look up $M_i$ in Table 1. If $M_i$ is
    equal to one of the value of $M$ in Table 1, take the
    corresponding ciphertext message $D_1,D_2,\ldots,D_q$
    and compute
    \begin{eqnarray*}
    H_1 & = & e_{K_1}(D_1),\\
    H_\ell & = & e_{K_1}(D_\ell\oplus H_{\ell-1}),\:\:\:(2\leq \ell\leq q), \mbox{ and }\\
    B & = & H_q.
    \end{eqnarray*}
    Now store $B$ together with $i$ in Table 2, which is
    sorted (or hashed) on the $B$ values.  Note that the
    same $B$ value may occur more than once.
\item Each entry in Table 2 consists of a value of $B$ and
    the corresponding key $i$, where $i$ is a candidate for
    $K_1$; as described above, each ($B$, $i$) pair is
    associated with a (($D_1, D_2, \ldots, D_q$), $M$) pair
    from Table 1 where $d_i(M) = A$. The remaining task is
    to search for possible values of $K_2$.

    For each of the $2^{56}$ candidates, $j$, for $K_2$,
    calculate what the value $B$ would be if $j$ had been
    used for $K_2$, i.e.\ $B_j = e_j(A)$. Now look up $B_j$
    in Table 2. For each appearance of $B_j$ (if any) the
    corresponding key $i$ from Table 2, along with key $j$,
    is a candidate for the desired pair of keys ($K_1$,
    $K_2$). Each such candidate key pair is then tested on
    at most two other message/MAC pairs. If this key pair
    gives the correct results then the target DES key pair
    ($K_1$, $K_2$) has been found and the task is complete.
\item If the algorithm does not succeed, then the process
    in steps 2--4 is repeated for a new value of $A$.
\end{enumerate}

\subsection{Impact on security}

The algorithm is, of course, very similar to that given in
section~\ref{van}.  As a result, the complexity considerations
are very similar too, with the following exceptions.
\begin{itemize}
\item Table 1 is now larger, containing a $q$-block message
    and a 64-bit MAC.  If, as above, we assume $q\leq 2^r$
    for some $r$, then Table 1 will contain at most
    $2^{r+3}n$ bytes.
\item Computing an entry in Table 2 will take up to $r+1$
    DES computations instead of a single DES computation.
\item Checking a candidate key pair will take up to $r+2$
    DES computations.
\end{itemize}
However, as long as $r$ is not too large, say $r\leq 2^{10}$
then even if $n$ is as large as $2^{40}$, Table 1 will contain
at most $2^{53}$ bytes.  Since the number of entries in Table 2
is much less than in Table 1, and similarly the number of
candidate key pairs is much less than $2^{56}$, the other two
differences do not affect the overall attack complexity.

Hence, bearing in mind the generalisations to the van
Oorschot-Wiener attack described above, if we have $2^t$ known
message/MAC pairs, i.e.\ $n=2^t$, and the message length is
bounded above by $2^r$, then the ANSI retail MAC can be broken
using $2^{120-t}$ DES computations and $O(2^{t+r})$ storage.
For example, if $n=2^{32}$, i.e.\ if we have as many as 4
billion known message/MAC pairs, then one of the DES key pairs
used can be discovered in $2^{88}$ DES computations.  The main
novel observation here is that the known message/MAC pairs do
not need to all have been generated using the same pair of
keys.

Hence if, for example, no more than $2^{30}$ message/MAC pairs
are available generated using a single key, the Preneel-van
Oorschot attack will simply not apply, whereas the attack
described above will.  That is, limiting the number of MACs
generated using a single pair of DES keys, whilst effective in
mitigating the Preneel-van Oorschot attack, does not protect
against the generalised van Oorschot-Wiener attack.

\section{Conclusions --- the future of 2-key triple
DES}  \label{conclusions}

The fact that the van Oorschot-Wiener attack can be used with
both plaintext/ciphertext pairs generated using a multiplicity
of keys and with partially known plaintext significantly
enlarges the set of scenarios in which the security of 2-key
triple DES is at risk.  Whilst obtaining $2^{32}$ known
plaintext-ciphertext pairs all generated using a single key
sounds like a tall order for an attacker, obtaining the same
number of only partially known plaintext/ciphertext pairs
possibly generated using a multiplicity of keys seems greatly
more plausible. This is why we suggest that the estimate of
80-bit security seems a very realistic estimate, and does not
leave much margin of safety.

In particular, the advice to change keys regularly does not
give the protection expected.  Of course, performing regular
key changes is good advice, but does not reduce the success
probability of the attack; it only limits the impact of a
successful attack.

80 bits of security does not seem very much today, given that
56 bits of security, as provided by single DES, was deemed very
risky 30 or more years ago. It would therefore seem prudent to
replace 2-key triple DES as soon as possible, either with the
3-key variant or with a more modern and more efficient
algorithm like AES.  Use of AES also allows the introduction of
256-bit keys, giving protection against possible attacks based
on quantum computing.

As a final remark we also observe that the observations in
section~\ref{ARMAC} also cast doubt on the future viability of
the ANSI retail MAC when used with DES.

\bibliography{Crypto}

\begin{thebibliography}{10}

\bibitem{K128}
American Bankers Association, Washington, DC.
\newblock {\em {ANSI} {X}9.19, Financial institution retail message
  authentication}, August 1986.

\bibitem{kn:107}
American National Standards Institute, New York.
\newblock {\em ANSI X3.92--1981, {D}ata {E}ncryption {A}lgorithm}, 1981.

\bibitem{K769}
American National Standards Institute, New York.
\newblock {\em ANSI X9.52--1998, Triple Data Encryption Algorithm --- Modes of
  operation}, 1998.

\bibitem{K774}
M.~Curtin and J.~Dolske.
\newblock A brute force search of {DES} keyspace.
\newblock \url{http://www.interhack.net/pubs/des-key-crack/}, November 1998.
\newblock [Online --- accessed 2nd February 2016].

\bibitem{K688}
W.~Diffie and M.~Hellman.
\newblock Exhaustive cryptanalysis of the {NBS} data encryption standard.
\newblock {\em IEEE Computer}, 10(6):74--84, June 1977.

\bibitem{K770}
{Electronic Frontier Foundation}.
\newblock {\em Cracking DES: Secrets of encryption research, wiretap politics
  and chip design}.
\newblock O'Reilly and Associates, 1998.

\bibitem{K777}
EMVCo.
\newblock {\em {Integrated Circuit Card Specifications for Payment Systems ---
  Book 2: Security and Key Management}}, November 2011.
\newblock Version 4.3.

\bibitem{K313}
International Organization for Standardization, Gen\`{e}ve, Switzerland.
\newblock {\em {ISO}/{IEC} 9797--1, {I}nformation technology --- {S}ecurity
  techniques --- {M}essage {A}uthentication {C}odes ({MAC}s) --- {P}art 1:
  {M}echanisms using a block cipher}, 1999.

\bibitem{K775}
International Organization for Standardization, Gen\`{e}ve, Switzerland.
\newblock {\em ISO 9564--1:2002, Banking --- {Personal Identification Number
  (PIN)} management and security --- {Part 1: Basic} principles and
  requirements for online {PIN} handling in {ATM} and {POS} systems}, 2nd
  edition, 2002.

\bibitem{K534}
International Organization for Standardization, Gen\`{e}ve, Switzerland.
\newblock {\em {ISO}/{IEC} 18033--3, {Information} technology --- {S}ecurity
  techniques --- {E}ncryption algorithms --- {P}art 3: Block ciphers}, 2005.

\bibitem{K702}
International Organization for Standardization, Gen\`{e}ve, Switzerland.
\newblock {\em {ISO}/{IEC} 18033-3:2010, {Information} technology ---
  {S}ecurity techniques --- {E}ncryption algorithms --- {P}art 3: Block
  ciphers}, 2nd edition, 2010.

\bibitem{K772}
International Organization for Standardization, Gen\`{e}ve, Switzerland.
\newblock {\em ISO/IEC JTC 1/SC 27 N13432, ISO/IEC JTC 1/SC 27 Standing
  Document No.\ 12 (SD12) on the Assessment of Cryptographic Techniques and Key
  Lengths, 4th edition}, May 2014.

\bibitem{K773}
International Organization for Standardization, Gen\`{e}ve, Switzerland.
\newblock {\em ISO/IEC JTC 1/SC 27 N14908, First edition of SC 27/WG 2 Standing
  Document 4 --- Analysis and status of cryptographic algorithms}, December
  2014.

\bibitem{K358}
L.~R. Knudsen and C.~J. Mitchell.
\newblock Analysis of 3gpp-{MAC} and two-key 3gpp-{MAC}.
\newblock {\em Discrete Applied Mathematics}, 128:181--191, 2003.

\bibitem{K292}
L.~R. Knudsen and B.~Preneel.
\newblock Mac{DES}: {MAC} algorithm based on {DES}.
\newblock {\em Electronics Letters}, {\bf 34}:871--873, 1998.

\bibitem{K780}
S.~Lucks.
\newblock Attacking triple encryption.
\newblock In S.~Vaudenay, editor, {\em Fast Software Encryption, 5th
  International Workshop, {FSE} '98, Paris, France, March 23--25, 1998,
  Proceedings}, volume 1372 of {\em Lecture Notes in Computer Science}, pages
  239--253. Springer-Verlag, Berlin, 1998.

\bibitem{K276}
A.~J. Menezes, P.~C. van Oorschot, and S.~A. Vanstone.
\newblock {\em Handbook of Applied Cryptography}.
\newblock CRC Press, Boca Raton, 1997.

\bibitem{K771}
R.~C. Merkle and M.~E. Hellman.
\newblock On the security of multiple encryption.
\newblock {\em Communications of the ACM}, 24(7):465--467, 1981.

\bibitem{K364}
C.~J. Mitchell.
\newblock Key recovery attack on {ANSI} retail {MAC}.
\newblock {\em Electronics Letters}, 39:361--362, 2003.

\bibitem{K363}
National Institute of Standards and Technology (NIST), Gaithersburg, MD.
\newblock {\em Federal Information Processing Standards Publication 46-3 (FIPS
  PUB 46-3): {D}ata {E}ncryption {S}tandard ({DES})}, October 1999.

\bibitem{K361}
National Institute of Standards and Technology (NIST), Gaithersburg, MD.
\newblock {\em Federal Information Processing Standards Publication 197 (FIPS
  PUB 197): Specification for the Advanced Encryption Standard (AES)}, November
  2001.

\bibitem{K767}
National Institute of Standards and Technology (NIST).
\newblock {\em {NIST} {S}pecial {P}ublication {800-67}, {R}ecommendation for
  the {Triple Data Encryption Algorithm (TDEA)} block cipher}, January 2012.
\newblock Revision 1.

\bibitem{NIST800-131A}
National Institute of Standards and Technology (NIST).
\newblock {\em {NIST} {S}pecial {P}ublication {800-131A}, {Transitions:
  Recommendations} for transitioning the use of cryptographic algorithms and
  key lengths}, November 2015.
\newblock Revision 1.

\bibitem{K776}
National Institute of Standards and Technology (NIST).
\newblock {\em {NIST} {S}pecial {P}ublication 800--57 Part 1, {R}ecommendation
  for {Key Management}}, January 2016.
\newblock Revision 4.

\bibitem{kn:108}
National Technical Information Service, Springfield, Va.
\newblock {\em National Bureau of Standards (NBS) Federal Information
  Processing Standards (FIPS) Publication 46---{D}ata {E}ncryption {S}tandard
  ({DES})}, April 1977.

\bibitem{K300}
B.~Preneel and P.~C. van Oorschot.
\newblock A key recovery attack on the {ANSI} {X}9.19 retail {MAC}.
\newblock {\em Electronics Letters}, {\bf 32}:1568--1569, 1996.

\bibitem{K350}
P.~C. van Oorschot and M.~J. Wiener.
\newblock A known plaintext attack on two-key triple encryption.
\newblock In I.~B. Damgard, editor, {\em Advances in Cryptology --- EUROCRYPT
  '90}, number 473 in Lecture Notes in Computer Science, pages 318--325.
  Springer-Verlag, Berlin, 1991.

\end{thebibliography}

\end{document}